\documentclass[12pt]{iopart}

\usepackage[T1]{fontenc}
\usepackage{iopams}
\usepackage{mathrsfs}
\usepackage{url}
\usepackage{collref}
\usepackage{graphicx}
\usepackage{enumerate}
\usepackage{eucal}
\usepackage{subfigure}
\usepackage{tensor}
\usepackage[perpage,symbol*]{footmisc}
\DefineFNsymbols{akreienb}{\dag\ddag\S\P}
\setfnsymbol{akreienb}

\newcommand{\mca}{\CMcal}

\newcommand{\msf}{\mathsf}

\newcommand{\mbb}{\mathbb}
\newcommand{\mrm}{\mathrm}

\newcommand{\tit}{\textit}

\newcommand{\pde}{\partial}

\newcommand{\pel}{\parallel}

\let\oldsqrt\sqrt
\def\sqrt{\mathpalette\DHLhksqrt}
\def\DHLhksqrt#1#2{
\setbox0=\hbox{$#1\oldsqrt{#2\,}$}\dimen0=\ht0
\advance\dimen0-0.2\ht0
\setbox2=\hbox{\vrule height\ht0 depth -\dimen0}
{\box0\lower0.4pt\box2}}

\begin{document}

\title{Modified general relativity as a model for quantum gravitational collapse}

\author{Andreas Kreienbuehl$^1$, Viqar Husain$^2$, and Sanjeev S. Seahra$^2$}
\address{$^1$ Institute for Theoretical Physics, Utrecht University,\\
3584 CE Utrecht, The Netherlands\\
$^2$ Department of Mathematics and Statistics, University of New Brunswick,\\
Fredericton, NB E3B 5A3, Canada}
\eads{\mailto{a.kreienbuehl@uu.nl}, \mailto{vhusain@unb.ca}, and \mailto{sseahra@unb.ca}}

\begin{abstract}
We study a class of Hamiltonian deformations of the massless Einstein-Klein-Gordon system in spherical symmetry for which the Dirac constraint algebra closes. The system may be regarded as providing effective equations for quantum gravitational collapse. Guided by the observation that scalar field fluxes do not follow metric null directions due to the deformation, we find that the equations take a simple form in characteristic coordinates. We analyse these equations by a unique combination of numerical methods and find that Choptuik's mass scaling law is modified by a mass gap as well as jagged oscillations. Furthermore, the results are universal with respect to different initial data profiles and robust under changes of the deformation.
\end{abstract}

%\maketitle

\section{Introduction}

Einstein's theory of general relativity (GR) predicts the existence of black holes. According to the Penrose-Hawking singularity theorems they can form when matter undergoes gravitational collapse. However, the singularities inside black holes are unwelcome from a physical point of view as they render spacetime incomplete. We have no knowledge about what happens near a singularity. This is a regime where quantum gravity effects are expected to come into play.

The canonical formulation of GR is one approach to formulating a theory of quantum gravity. It can be traced back to the work of Arnowitt, Deser, and Misner (ADM) \cite{arnowitt:62} who gave the first Hamiltonian formulation. This was used by Wheeler \cite{wheeler:68} and DeWitt \cite{dewitt:67} to formulate the quantization program. In practice, there are many technical and conceptual difficulties in this approach \cite{isham:91,kuchar:92,isham:93}, and so far no complete theory is available.

In the absence of a full theory it is important to ask whether there are relevant physical situations with symmetries where the canonical quantization can be completed. Technically the easiest is that of a homogeneous cosmology, in which case GR reduces to a system with finitely many degrees of freedom. Such reductions have been studied since the 1970s using the Wheeler-DeWitt approach \cite{misner:72}, and again in the recent past using loop quantum gravity (LQG) methods \cite{bojowald:08b,ashtekar:09}.

The next level of problem, beyond quantum mechanics, is dimensional reduction to a tractable field theory. Two important problems of physical interest that fall into this category are that of a cosmology with inhomogeneities and that of the spherically symmetric gravitational collapse. Our interest in this paper is in the second problem. We address the question of how to formulate deformations of Einstein's equations in spherical symmetry and discuss how they can be interpreted as providing ``effective'' quantum gravity corrections using some input from LQG. 

Classical spherically symmetric collapse equations have been extensively researched. Using Schwarzschild coordinates \cite{choptuik:86}, Christodoulou showed \cite{christodoulou:84} that there are two classes of initial data for the collapse. For one of them (``weak data'') matter bounces at the centre of the coordinate system and evolves back towards radial infinity. The other class of data, however, gives rise to a black hole. This was numerically confirmed not only by Choptuik \cite{choptuik:86} but also by Goldwirth and Piran \cite{goldwirth:87}, who used infalling null coordinates \cite{christodoulou:86}. For a comparison of the two approaches see \cite{choptuik:92}. The next important development in the history of classical spherically symmetric gravitational collapse was by Choptuik \cite{choptuik:93}. Among other things, he found numerically a scaling law which relates the amplitude of the scalar field to the mass of the black hole. This was later reproduced by Garfinkle \cite{garfinkle:94} based on collapse equations in double null coordinates. Such a form of the equations was also used in \cite{husain:01,birukou:02,husain:03a} to find a similar scaling law with a nonzero cosmological constant and in higher dimensions.

The investigation of quantum corrected spherically symmetric collapse was the natural next step to take. In \cite{husain:09} it was shown that certain corrections motivated by the LQG program introduce a mass gap in the mass scaling law. This indicates that zero mass black holes are no longer contained in the solution space. Using similar corrections in Painlev\'e-Gullstrand coordinates \cite{husain:05b}, Ziprick and Kunstatter \cite{ziprick:09a,ziprick:10} also found the mass gap, and importantly, were able to observe the evolution beyond the horizon. 

In none of the works mentioned in the previous paragraph have the modifications been directly introduced in the context of a Hamiltonian formulation. Rather, the corrections were inserted into the evolution equations or a covariant action (in \cite{ziprick:10} this is the so-called dilaton action for GR). Also, at least in \cite{husain:09,ziprick:09a}, it has not been checked whether the modified equations are consistent with GR's symmetry group, that is the diffeomorphisms. It is therefore interesting to ask if it is possible to arrive at effective equations representing quantum gravity corrections from a Hamiltonian formulation and that are such that the Dirac algebra closes. This forms part of the motivation for the present project. 

Following work on loop quantum cosmology by Bojowald and his col\-la\-bo\-rat\-ors \cite{bojowald:06,bojowald:07,bojowald:08a}, where a class of anomaly-free effective constraints are presented, Reyes \cite{reyes:09} obtained a closed constraint algebra in spherical symmetry using the con\-nec\-tion-triad variables of LQG. However, the resulting equations are sufficiently complicated as to make a numerical investigation quite difficult. 

The equations we derive and analyse in this paper follow from the same general ideas but are obtained directly in the ADM formulation. The main obstacle faced with the connection-triad variables is overcome to yield a remarkably simple form of the equations that can be compared directly to the classical double null formulation. Along the way we note a useful physical insight crucial for the form of the equations we obtain: the null coordinates of the metric do not coincide with the characteristic lines of the scalar field due to the deformation. This is what may be expected to be necessary for singularity avoidance, since it signals a violation of the dominant energy condition. 

The paper is structured as follows. In the next section we present the canonical theory and introduce a class of deformations that are to model quantum corrections. In Section \ref{s:11} we derive the corresponding canonical equations of motion for the Schwarzschild gauge. The transformation to characteristic coordinates, which yields the simple equations, is given in Section \ref{s:12}. In Section \ref{s:51} we choose a specific function that deforms GR. The numerical scheme to solve the resulting equations is defined in Section \ref{s:52}. The definition of black holes requires some commentary, which is provided in Section \ref{s:53}. We proceed in Section \ref{s:54} by posing the questions that we want to answer in the remainder of the paper. Finally, Section \ref{s:100} contains the results and Section \ref{s:56} some concluding remarks.

\section{Canonical formulation and its deformation}

In metric ADM variables the line element for a spherically symmetric spacetime can be defined by 
\begin{equation}\label{e:517}
	ds^2\equiv-[N^2-(N^rA)^2]dt^2+2N^rA^2dtdr+A^2dr^2+B^2d\Omega^2,
\end{equation}
where $N$ is the lapse function and $N^r$ the radial component of the shift vector. If, to this metric, we minimally couple a massless scalar field with matter Lagrangian density $-\sqrt{-|g|}g^{\mu\nu}\pde_{\mu}\phi\pde_{\nu}\phi/(8\pi)$, we get the canonical action\footnote{The boundary terms can be ignored here.}
\begin{equation}\label{e:523}
	S=\int_{t_0}^{t_1}\int_0^{\infty}(p_A\dot{A}+p_B\dot{B}+p_{\phi}\dot{\phi}-\mca{H})\;dr\;dt.
\end{equation}
The Poisson brackets are
\begin{equation*}
	\{A(t,r),p_A(t,s)\}=\{B(t,r),p_B(t,s)\}=\{\phi(t,r),p_{\phi}(t,s)\}=\delta(r,s)
\end{equation*}
and the total Hamiltonian density $\mca{H}\equiv N\mca{H}_{\perp}+N^r\mca{H}_{\pel r}$ consists of the Hamiltonian constraint density\footnote{We set $G_{\mrm{N}}\equiv1$ and $c\equiv1$.}
\begin{equation}\label{e:51}
	\eqalign{
		\mca{H}_{\perp}&\equiv\frac{p_A}{2B^2}(p_AA-2p_BB)-\frac{1}{2A^2}[A'B^{2\prime}-A(2BB''+B'^2)+A^3]\cr
		&\quad+\frac{p_{\phi}^2}{2AB^2}+\frac{B^2\phi'^2}{2A}
	}
\end{equation}
and the diffeomorphism constraint density 
\begin{equation*}	
	\mca{H}_{\pel r}\equiv-p_A'A+p_BB'+p_{\phi}\phi'.
\end{equation*}

From this well known setup \cite{berger:72,unruh:76,kuchar:94} we can introduce a classical deformation that is designed to describe a certain type of quantum effect. As noted above, the way we arrive at the correction is based on the ideas of Bojowald and his collaborators \cite{bojowald:06,bojowald:07,bojowald:08a}, which were also applied by Reyes \cite{reyes:09}. One of the differences here is that we are not using the connection-triad variables \cite{ashtekar:87,barbero:95,immirzi:97}. 

The insight motivating the deformation is that the inverse powers of $A$ and $B$ in \eref{e:51} can be conveniently quantized using the Thiemann prescription for the inverse triad operator \cite{thiemann:98a}.\footnote{We note that this operator is not the only source for quantum corrections; the other is the curvature written using holonomy operators. It is not known how to get a deformed, anomaly-free effective system that includes this correction in general, but some special cases have been studied \cite{reyes:09,bojowald:08c,bojowald:09}.} Doing so, we can turn $H_{\perp}\equiv\int_0^{\infty}N\mca{H}_{\perp}\;dr$ into an operator $\hat{H}_{\perp}$. If we proceed by computing the expectation value of $\hat{H}_{\perp}$ for certain classes of states\footnote{See \cite{husain:07a} for the case of cosmological models.}, we get an effective $H_{\perp}^{\mrm{e}}$. The afore cited work by Bojowald and others suggests that it can be of the form $H_{\perp}^{\mrm{e}}\equiv\int_0^{\infty}N\mca{H}_{\perp}^{\mrm{e}}\;dr$, where the effective Hamiltonian constraint density is
\begin{equation}\label{e:52}
	\mca{H}_{\perp}^{\mrm{e}}\equiv\sum_{i=1}^3Q_{(i)}\mca{H}_{\perp}^{(i)}
\end{equation}
with
\begin{equation*}
	\eqalign{
		&\mca{H}_{\perp}^{(1)}\equiv\frac{p_A}{2B^2}(p_AA-2p_BB)-\frac{1}{2A^2}[A'B^{2\prime}-A(2BB''+B'^2)+A^3],\cr
		&\mca{H}_{\perp}^{(2)}\equiv\frac{p_{\phi}^2}{2AB^2},\qquad \mca{H}_{\perp}^{(3)}\equiv\frac{B^2\phi'^2}{2A}.
	}
\end{equation*}
In this definition, the $Q_{(i)}$ depend only on $B$ and the $\mca{H}_{\perp}^{(i)}$ describe, in increasing order of $i$, the extrinsic curvature plus the Ricci scalar of constant parameter time $t$ hypersurfaces, the kinetic energy of the scalar field, and its gradient energy. The form of \eref{e:52} implies that the deformation is entirely contained in the functions $Q_{(i)}$. Moreover, by restricting the $Q_{(i)}$ to depend only on $B$, modifications due to terms involving inverse powers of $A$ are disregarded. Lastly, the term stemming from the extrinsic curvature and the one from the Ricci scalar are treated equally even though their overall dependence on $B$ is different. 

In spite of these concerns, there is a very good reason to investigate the physical consequences of \eref{e:52}. Namely, if we impose the condition
\begin{equation}\label{e:516}
	Q_{(1)}^2=Q_{(2)}Q_{(3)}
\end{equation}
it can be shown (as in \cite{reyes:09}) that the Dirac constraint algebra closes. There are no anomalies if \eref{e:516} is true. This has far reaching consequences. Among them is that all degrees of freedom remain to be captured by the given canonical variables. Furthermore, it implies that energy is conserved, which is attractive from a physical point of view.

\section{Equations in Schwarzschild coordinates\label{s:11}}

To get an idea of the effect of the quantum corrections introduced in the previous section, we implement the second class gauges $\mca{G}_{\pel r}\equiv r-B=0$ and $\mca{G}_{\perp}\equiv p_A=0$ in the given order to freeze out some of the degrees of freedom \cite{dirac:64,hanson:76,henneaux:92}. Concretely, the consistency condition $\dot{\mca{G}}_{\pel r}=0$ gives $N^r=Np_A/r$ and the equation $\mca{H}_{\pel r}=0$ yields 
\begin{equation}\label{e:518}
	p_B=p_A'A-p_{\phi}\phi'.
\end{equation}
Note that the gauge choice $\mca{G}_{\pel r}=0$ renders the $Q_{(i)}$ nondynamical by turning them into functions that solely depend on $r$. Proceeding with the gauge fixing, the choice of $\mca{G}_{\perp}$ forces $N^r$ to vanish, which not only turns the line element \eref{e:517} into the familiar form
\begin{equation}\label{e:546}
	ds^2=-N^2dt^2+A^2dr^2+r^2d\Omega^2
\end{equation}
but also, from \eref{e:518} and the definition of $p_B$, implies the trivial equation
\begin{equation}\label{e:519}
	\dot{A}=\frac{Q_{(1)}Np_{\phi}\phi'}{r}.
\end{equation}
The consistency condition $\dot{\mca{G}}_{\perp}=0$ leads to
\begin{equation}\label{e:520}
	\frac{N'}{N}=\frac{A'}{A}+\frac{A^2-1}{r}-\frac{Q_{(1)}'}{Q_{(1)}}
\end{equation}
and solving $\mca{H}_{\perp}^{\mrm{e}}=0$ for $A$ gives
\begin{equation}\label{e:521}	
	\frac{A'}{A}=\frac{1-A^2}{2r}+\frac{1}{2r^3Q_{(1)}}(Q_{(2)}p_{\phi}^2+r^4Q_{(3)}\phi'^2).
\end{equation}
Finally, Hamilton's form of the scalar field equation is given by
\begin{equation}\label{e:522}
	\dot{\phi}=\frac{Q_{(2)}Np_{\phi}}{r^2A},\qquad\dot{p}_{\phi}=\left(\frac{r^2Q_{(3)}N\phi'}{A}\right)'.
\end{equation}
 
If we set all $Q_{(i)}$ equal to $1$, the deformation is switched off. This allows us to recover in (\ref{e:519})-(\ref{e:522}) the familiar collapse equations in Schwarzschild coordinates. This is the form of the equations that has been heavily investigated analytically and numerically (see \cite{gundlach:07} for an extensive list of references). 

However, to reproduce Choptuik's numerically found results in these coordinates is a rather complex task. The conceptual reason is that near $r=0$, where black holes are expected to form, a numerical code has to allow for a very high resolution but as we move away from the origin there is no need for the same accuracy. Trying to keep the computing time low, Choptuik decided to use the adaptive mesh refinement algorithm of Berger and Oliger \cite{berger:84}. Unfortunately, and despite its beauty, the method is rather involved so that the conceptual problem is effectively replaced by a practical one. Because of this we seek a different form of the deformed equations, one that is more user-friendly.

\section{Equations in characteristic coordinates\label{s:12}}

Christodoulou \cite{christodoulou:86} showed that Einstein's field equations for the model defined by \eref{e:523} can be given in a very compact form if infalling null coordinates $(u,r)$ are used. Some years later, Garfinkle \cite{garfinkle:94} derived the same equations in double null coordinates $(u,v)$. Namely, in terms of a line element of the form
\begin{equation*}
	ds^2\equiv-W^2dudv+r^2d\Omega^2,
\end{equation*}
Einstein's field equations are given by \cite{christodoulou:93}
\begin{equation}\label{e:528}
	\eqalign{
		&W^2=-2\pde_u\pde_vr^2,\cr
		&2\pde_ur\frac{\pde_uW}{W}=\pde_u^2r+r(\pde_u\phi)^2,\cr
		&2\pde_vr\frac{\pde_vW}{W}=\pde_v^2r+r(\pde_v\phi)^2,\cr
		&\pde_ur\pde_v\phi+\pde_vr\pde_u\phi=-r\pde_u\pde_v\phi,
	}
\end{equation}
and due to the Bianchi identities the second and the third equation are redundant \cite{christodoulou:86}. Dropping the second, Garfinkle used the reparametrization $W^2\equiv2\pde_vrF$ to arrive at the evolution equations
\begin{equation}\label{e:529}
	\pde_ur=-\frac{f}{2},\qquad\pde_u\Phi=\frac{1}{2r}(f-F)(\phi-\Phi),
\end{equation}
where
\begin{equation}\label{e:540}
	\eqalign{
		&\pde_v\phi=-\frac{\pde_vr}{r}(\phi-\Phi),\cr
		&\pde_vF=\frac{\pde_vr}{r}F(\phi-\Phi)^2,\cr
		&\pde_vf=-\frac{\pde_vr}{r}(f-F).
	}
\end{equation}
These equations are of the same compact form as those presented by Christ\-odou\-lou for infalling null coordinates.

Equations \eref{e:529} and \eref{e:540} have several advantages compared to those in Schwarzschild coordinates (\ref{e:519})-(\ref{e:522}) (with the deformation switched off). The most important is that in double null coordinates $(u,v)$ we have $\pde_ur<0$, which follows from the first equation in \eref{e:528} \cite{christodoulou:93}. This implies that the computational grid becomes smaller as $r=0$ is approached so that a higher accuracy can be naturally obtained by inserting additional points into constant $u$ slices \cite{garfinkle:94,husain:01,birukou:02,husain:03a,husain:09}. An adaptive mesh refinement algorithm in the spirit of that used by Choptuik is no longer necessary. We therefore intend to transform our deformed equations to double null coordinates. However, as we will see, this is not enough to obtain simple equations due to a null characteristic mismatch. 

To simplify the following expressions, we introduce the functions $Q\equiv(Q_{(2)}/Q_{(1)})^{1/2}$ and $s\equiv A/(Q_{(1)}N)$ as well as the matter variables 
\begin{equation*}
	\chi\equiv(s\dot{\phi}-\phi')/Q,\qquad\psi\equiv(s\dot{\phi}+\phi')/Q.
\end{equation*}
This allows us to write the trivially solved Equation \eref{e:519} in the form
\begin{equation}\label{e:535}
	\frac{s\dot{A}}{A}=-\frac{r}{4}(\chi^2-\psi^2)
\end{equation}
and the remaining Equations (\ref{e:520})-(\ref{e:522}) as
\begin{equation}\label{e:514}
	\eqalign{
		&\frac{s'}{s}=\frac{1-A^2}{r},\qquad\frac{A'}{A}=\frac{1-A^2}{2r}+\frac{r}{4}(\chi^2+\psi^2),\cr
		&s\dot{\chi}+\chi'=-\ln(r/s)'\chi+\ln(r/Q)'\psi,\cr
		&s\dot{\psi}-\psi'=-\ln(r/Q)'\chi+\ln(r/s)'\psi.
	}
\end{equation}
The last two equations imply that we can denote by $c_{\pm}^{\mu}\pde_{\mu}\equiv s\pde_t\pm\pde_r$ the characteristic directions \cite{strikwerda:04} of the scalar field equation. By definition, the variable $\chi$ describes the change of $\phi$ along $c_-$ with speed $-s$, and $\psi$ does so along $c_+$ with speed $s$. Therefore, the last two equations in \eref{e:514} tell us how a consecutive change of $\phi$ along $c_-$ and $c_+$, and vice versa, looks like. 

Note, however, that \tit{the characteristics do not correspond to the null directions} $Q_{(1)}s\pde_t\pm\pde_r$ \tit{of the metric} \eref{e:546}. In fact, we have 
\begin{equation}\label{e:531}
	ds(c_{\pm},c_{\pm})^2=-\frac{1-Q_{(1)}^2}{Q_{(1)}^2}A^2,
\end{equation}
which vanishes if and only if $Q_{(1)}=1$. This means that the fluxes of the massless scalar field do not follow null directions so that the theory under investigation is not GR.

The important implication of \eref{e:531} is that only if $Q_{(1)}^2\leq1$ are we dealing with matter of which the fluxes are timelike or null. If $Q_{(1)}^2>1$ the scalar field propagates along spacelike directions, giving a violation of the dominant energy condition (see also \cite{ziprick:10}). In this case we can hypothesise that the formation of singularities can be avoided. 

What kind of matter are we dealing with? In this context, it is interesting to note that there is no scalar field potential that can be added to the Lagrangian matter density to reproduce \eref{e:535} and \eref{e:514} by means of Einstein's equations. This follows from the fact that a potential not only gives a contribution to the scalar field equation but also to the consistency condition and the constraint equation. However, in the given variables the latter two are effectively unaffected by the quantum corrections ($Q$ only appears in the last two equations in \eref{e:514}).

The equations for $\chi$ and $\psi$ in \eref{e:514} suggest that we introduce coordinates $u$ and $v$ adapted to the characteristics $c_{\pm}$. We therefore set
\begin{equation}\label{e:532}
	D_u\equiv-s\pde_t+\pde_r,\qquad D_v\equiv s\pde_t+\pde_r,
\end{equation}
where $D_u\equiv(\pde_ur)^{-1}\pde_u$ and $D_v\equiv(\pde_vr)^{-1}\pde_v$. This gives 
\begin{equation*}
	\chi=-D_u\phi/Q,\qquad\psi=D_v\phi/Q.
\end{equation*}
That is, since $\pde_ur<0$ and $\pde_vr>0$ (see below), the coordinates $u$ and $v$ respectively parametrize future pointing infalling and outgoing characteristics if $Q_{(1)}^2<1$. For \eref{e:532} we chose $\pde_ut\equiv-s\pde_ur$ and $\pde_vt\equiv s\pde_vr$, which implies the coordinate relation
\begin{equation}\label{e:533}
	-W^2dudv=-Q_{(1)}^2N^2dt^2+A^2dr^2
\end{equation}
if the function $W$ is parametrized by
\begin{equation}\label{e:534}
	-\frac{W^2}{4\pde_ur\pde_vr}\equiv A^2.
\end{equation}
Since the characteristics $c_{\pm}$ to which we adapted the coordinates $u$ and $v$ do not agree with the null lines of the physical relevant line element \eref{e:546}, we of course cannot expect $u$ and $v$ to correspond to the double null coordinates of this metric. As is shown by \eref{e:533}, $u$ and $v$ correspond to the double null coordinates of that metric, for which the null directions agree with the characteristics. 

If we factor out $A^2$ in \eref{e:533}, we can show from \eref{e:534} that $s'/s=-\pde_u\pde_vr/(\pde_ur\pde_vr)$. This, together with the first equation in \eref{e:514}, implies
\begin{equation}\label{e:537}
	W^2=-2\pde_u\pde_vr^2.
\end{equation}
A comparison with \eref{e:528} shows that the deformation has no impact on this equation. From \eref{e:537} it follows that we again have $\pde_ur<0$ \cite{christodoulou:93} and since \eref{e:534} implies that $\pde_ur\pde_vr<0$ we get $\pde_vr>0$, as desired. Constructing linear combinations $s\dot{A}\pm A'$ from \eref{e:535} and \eref{e:514}, Equation \eref{e:534} gives
\begin{equation}\label{e:536}
	\eqalign{
		2\pde_ur\frac{\pde_uW}{W}=\pde_u^2r+r(\pde_u\phi/Q)^2,\cr
		2\pde_vr\frac{\pde_vW}{W}=\pde_v^2r+r(\pde_v\phi/Q)^2.
	}
\end{equation}
Since we are no longer dealing with Einstein's field equations, the Bianchi identities cannot be used to show a redundancy in \eref{e:536}. Luckily, the trivially solved Equation \eref{e:535} can be used here. Note that the equations in \eref{e:536} do not differ very much from the corresponding ones in \eref{e:528}. Finally, the equations for $\chi$ and $\psi$ in \eref{e:514} give the scalar field equation
\begin{equation}\label{e:515}
	\pde_u(r/Q)\pde_v\phi+\pde_v(r/Q)\pde_u\phi=-(r/Q)\pde_u\pde_v\phi
\end{equation}
and it is this equation, in which the quantum corrections are most prominent.\footnote{It is also apparent that one could obtain a modified dispersion relation from this equation, which is a topic of much recent interest.}

As expected, the speed $s$ does not appear in any of (\ref{e:537})-(\ref{e:515}). It cannot be ``seen'' along the characteristics. Therefore, the only quantum corrections left are encoded in $Q$. This implies that we can switch them off not only by setting $Q_{(1)}$, $Q_{(2)}$, and therefore $Q_{(3)}$ equal to $1$ but also by means of the relation $Q_{(1)}=Q_{(2)}=Q_{(3)}$. This is not very surprising since, as we have seen, the gauge $\mca{G}_{\perp}$ turns the $Q_{(i)}$ into nondynamical functions and if they are the same, \eref{e:523} and \eref{e:52} show that we can effectively remove them by reparametrizing the lapse function. The class of quantum corrections where all $Q_{(i)}$ agree are therefore uninteresting from a physical point of view.

To arrive at a form of the evolution equations that resembles the one given in \eref{e:529} and \eref{e:540} we set $W^2\equiv2\pde_vrF$. Dropping the first equation in \eref{e:536}, we can focus on the consistency and the constraint equation
\begin{equation}\label{e:538}
	\pde_vrF=-\pde_u\pde_vr^2,\qquad\pde_vr\frac{\pde_vF}{F}=r(\pde_v\phi/Q)^2,
\end{equation}
respectively, as well as on \eref{e:515}. The key now is to realize that the relevant radial coordinate in \eref{e:538} is $r$, whereas in \eref{e:515} it is $r/Q$. To remove this asymmetry we define
\begin{equation}\label{e:3434}
	q\equiv\frac{r}{Q}, 
\end{equation}
which can be seen as the result of exchanging $Q_{(2)}$ with $r^2Q_{(2)}$. The function $q$ behaves like $Q$ in the sense that it is nondynamical with respect to the Schwarzschild coordinates $(t,r)$. In terms of the characteristic coordinates $(u,v)$ its dynamic is implicitly determined by that of $r$. Finally, since $\dot{q}=0$ implies $D_uq=D_vq=q'$, we can arrive at the sought after evolution equations. They are 
\begin{equation}\label{e:550}
	\pde_ur=-\frac{f}{2},\qquad\pde_u\Phi=\frac{1}{2r}\left[\left(1-\frac{rq''}{q'}\right)f-F\right](\phi-\Phi),
\end{equation}
where
\begin{equation}\label{e:551}
	\eqalign{
		&\pde_v\phi=-\frac{\pde_vrq'}{q}(\phi-\Phi),\cr
		&\pde_vF=\frac{\pde_vrq'^2}{r}F(\phi-\Phi)^2,\cr
		&\pde_vf=-\frac{\pde_vr}{r}(f-F).
	}
\end{equation}
A comparison with \eref{e:529} and \eref{e:540} shows that the deformation is elegantly captured by relatively minimal modifications. 

It is now straightforward to produce an evolution scheme for these equations. This requires only one additional input, which is the selection of the function $q$ such that it gives modifications to the classical behavior in regions of sufficiently large energy density. 

\section{\label{s:51}Modifications}

The definition of $\mca{H}_{\perp}^{\mrm{e}}$ in \eref{e:52} suggests that the functions $Q_{(i)}(B)$ modify the factors $1/B^2$. Namely, according to \cite{husain:08,husain:09}, such modifications can be explicitly constructed from the \tit{expectation value} of the operator $\widehat{1/B}$ on the so-called polymer Hilbert space \cite{halvorson:03,ashtekar:03a}. Originally, this method was motivated by the use of the polymer quantization in the context of LQG and loop quantum cosmology, and the result that we want to take serious is that it can remove the functional singularities of $1/B^2$. That is, $1/B^2$ can be ``regularized'' on the real line and we will choose the $Q_{(i)}$ to do precisely this. In this sense we can interpret these functions as describing effective quantum gravity corrections. 

Since the gradient energy $\mca{H}_{\perp}^{(3)}$ is not inversely proportional to $B^2$, we can set $Q_{(3)}\equiv1$. The condition \eref{e:516} yields then $Q_{(1)}^2=Q_{(2)}$ and we are left with one independent $Q_{(i)}$. If we now go back to the gauge fixed sector where $B=r$, the function $q$ from \eref{e:3434} becomes
\begin{equation*}
	q=\frac{r}{\sqrt{Q_{(1)}}}.
\end{equation*}
Because of \eref{e:531} this means that the characteristics $u,v\in\mbb{R}$ of the scalar field equation are timelike wherever $q>r$ and spacelike for $q<r$.

Given our intention to think of $Q_{(1)}$ as a means to regularize $1/B^2=1/r^2$, we require $1/q<\infty$ for $r\to0$. We also demand that the impact of $q$ vanishes in the asymptotically flat region, where energy densities and, thus, modifications due to effective quantum gravity corrections should be negligible. The remaining conditions on $q$ are that it is strictly increasing and positive as, otherwise, either the term $q''/q'$ in \eref{e:550} or the term $q'/q$ in \eref{e:551} lead to singular evolution equations. A valid choice is therefore
\begin{equation}\label{e:604}
	q\equiv\sqrt{\lambda^2+r^2},
\end{equation}
where we can use the length scale $\lambda$ to give a more precise definition of the condition on $q$ in the asymptotic region. Namely, we require $|r-q|\ll\lambda$ for $r\gg\lambda$. Note that \eref{e:604} implies that the characteristics $u,v\in\mbb{R}$ are everywhere timelike.

\begin{figure}
\begin{center}
	\includegraphics[scale=1]{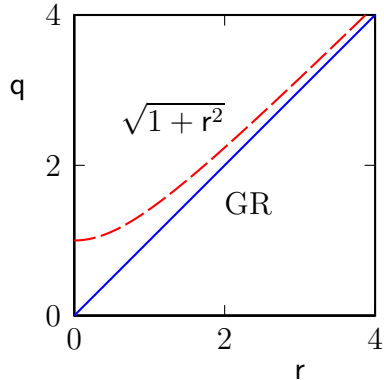}
	\caption[The function $\msf{q}=\sqrt{1+\msf{r}^2}$.]{\label{f:8}The function $\msf{q}=\sqrt{1+\msf{r}^2}$ modifies the GR curve $\msf{q}=\msf{r}$. Their difference vanishes in the limit $\msf{r}\to\infty$.}
\end{center}
\end{figure}

Having specified $q$, we can introduce dimensionless variables $\msf{u}$, $\msf{v}$, $\msf{r}$, and $\msf{q}$ by
\begin{equation*}
	\{u,v,r,q\}\longmapsto\{\lambda\msf{u},\lambda\msf{v},\lambda\msf{r},\lambda\msf{q}\}.
\end{equation*}
The equations of motion \eref{e:550} and \eref{e:551} are now dimensionless. The function $\msf{q}$ is now (see Figure \ref{f:8})
\begin{equation}\label{e:666}
	\msf{q}=\sqrt{1+\msf{r}^2}
\end{equation}
and we can move on to the numerics.

\section{\label{s:52}Numerical scheme}

The boundary conditions are $\msf{r}(\msf{u},\msf{u})=0$ \cite{christodoulou:93} and that $\Phi$ can be expanded in a series in $\msf{v}$ in a neighborhood of $\msf{v}=\msf{u}$. That is, we assume the existence of an expansion
\begin{equation}\label{e:605}
	\Phi(\msf{u},\msf{v})=\Phi(\msf{u},\msf{u})+(\pde_{\msf{v}}\Phi)(\msf{u},\msf{u})(\msf{v}-\msf{u})+\mrm{O}((\msf{v}-\msf{u})^2),
\end{equation}
which was also done in \cite{garfinkle:94}. The boundary conditions imply
\begin{equation*}
	F(\msf{u},\msf{u})=2(\pde_{\msf{v}}\msf{r})(\msf{u},\msf{u})=f(\msf{u},\msf{u}),\qquad(\pde_{\msf{v}}F)(\msf{u},\msf{u})=0=(\pde_{\msf{v}}f)(\msf{u},\msf{u}),
\end{equation*}
for the metric function $F$ and the auxiliary field $f$. The first relation is a consequence of the field equation $\pde_{\msf{v}}\msf{r}F=-\pde_{\msf{u}}\pde_{\msf{v}}\msf{r}^2$ together with $(\pde_{\msf{u}}\msf{r})(\msf{u},\msf{u})=-(\pde_{\msf{v}}\msf{r})(\msf{u},\msf{u})$. The second relation is only given as it is used in the implementation of the $\msf{v}$-integrator specified below. Regarding the boundary conditions for the matter fields, we have to have $\phi(\msf{u},\msf{u})=\Phi(\msf{u},\msf{u})$ in order to get evolution equations that are regular at $\msf{r}=0$. Finally, we need to know $(\pde_{\msf{v}}\phi)(\msf{u},\msf{u})$ and $(\pde_{\msf{u}}\Phi)(\msf{u},\msf{u})$ as they enter the setup of the $\msf{v}$-integrator and the $\msf{u}$-integrator, respectively (see below). They both depend on the choice of $\msf{q}$ and for \eref{e:666} we have
\begin{equation*}
	(\pde_{\msf{v}}\phi)(\msf{u},\msf{u})=0,\qquad(\pde_{\msf{u}}\Phi)(\msf{u},\msf{u})=(\pde_{\msf{v}}\Phi)(\msf{u},\msf{u}).
\end{equation*}

The initial data for $\msf{r}$ is specified by
\begin{equation}\label{e:610}
	\msf{r}(0,\msf{v})=\frac{\msf{v}}{2},
\end{equation}
which implies $F(0,0)=1$ \cite{christodoulou:93}. While the boundary condition $\msf{r}(\msf{u},\msf{u})=0$ removes the possibility of a reparametrization $\msf{u}\mapsto\tilde{\msf{u}}(\msf{u})$, \eref{e:610} makes $\msf{v}\mapsto\tilde{\msf{v}}(\msf{v})$ impossible. The remaining initial data is for the matter sector and is specified by a choice of $\phi$. Guided by \cite{choptuik:93} and others, we consider the profiles
\begin{eqnarray}
	\phi(0,\msf{v})&=&\phi_0\frac{\msf{v}^3}{1+\msf{v}^3}\exp\left(-\frac{(\msf{v}-\msf{v}_0)^2}{\msf{w}^2}\right),\label{e:609}\\
	\phi(0,\msf{v})&=&\phi_0\tanh\left(\frac{\msf{v}-\msf{v}_0}{\msf{w}}\right).\label{e:611}
\end{eqnarray}
The dependence on $\msf{v}$ in \eref{e:609} differs from that in the mentioned literature, where it is either $\phi_0\msf{v}^2\exp(\ldots)$ or $\phi_0\msf{v}^3\exp(\ldots)$. The factor $\msf{v}^3/(1+\msf{v}^3)$ allows us to choose $\msf{v}_0\gg1$ and, thus, to set up the initial data in the regime where the impact of $\msf{q}$ is minimal. That is, we do not have to worry about a black hole being already present on the slice $\msf{u}=0$. However, since $\msf{v}^3/(1+\msf{v}^3)=1+\mrm{O}(\msf{v}^{-3})$ for $\msf{v}\to\infty$, it seems likely that this factor disallows us to find a critical $\msf{v}_0^*\gg1$ for which Choptuik's mass scaling law applies (see Section \ref{s:54}). 

Our interest in the remainder of this paper is mainly in the initial data profile given in \eref{e:609}. The profile in \eref{e:611} is implemented in only one simulation and for the completeness of the analysis. It is important to note that \eref{e:609} is consistent with the boundary condition $\phi(\msf{u},\msf{u})=\Phi(\msf{u},\msf{u})$ since it implies $(\pde_{\msf{v}}^2\phi)(0,0)=0$. The latter is necessary because we have to solve $\Phi=\phi+2\msf{q}\pde_{\msf{v}}\phi/\msf{q}'$ for $\Phi$ on the initial slice, and the $\msf{q}$ in \eref{e:666} is such that $\msf{q}'(0)=0$. The initial data in \eref{e:611} satisfies the condition $(\pde_{\msf{v}}^2\phi)(0,0)=0$ the better the larger $\msf{v}_0$ is. Finally, given the form of $\phi(0,\msf{v})$ in \eref{e:609} and the above comments on $\msf{v}_0$, the parameters of interest are $\phi_0$ and $\msf{w}$, that is the amplitude and the width of the Gaussian-like wave packet, respectively.

\begin{figure}
\begin{center}
	\includegraphics{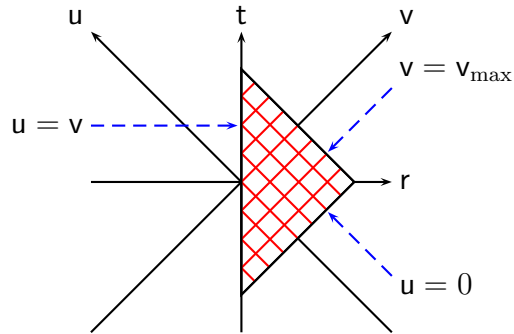}
	\caption[The numerical domain of integration.]{\label{f:5}The numerical domain of integration. For $\msf{q}\neq\msf{r}$ the lines $\msf{u},\msf{v}\in\mbb{R}$ do not coincide with the null directions of the metric.}
\end{center}
\end{figure}

To solve \eref{e:550} numerically, we need a $\msf{u}$-integrator. Since we are effectively dealing with a modified wave equation, we choose the Richtmyer two-step Lax-Wendroff method \cite{hoffman:01}. It is particularly well adapted to the computational domain, of which an illustration is given in Figure \ref{f:5}.
Note that the domain becomes smaller as $\msf{u}$ increases. Garfinkle \cite{garfinkle:94} made use of this by injecting additional points into slices of constant coordinate time $\msf{u}$, whenever one half of all the points disappeared beyond the boundary at $\msf{r}=0$. This is a simple method to achieve an increase in the resolution as the evolution progresses. However, we do not use this method and \tit{derive all results on the basis of a uni-grid code}. Finally, we choose the well known second-order Runge-Kutta method as the $\msf{v}$-integrator to solve \eref{e:551}. 

The resulting code is second-order accurate \cite{choptuik:86,choptuik:91,pretorius:02,pretorius:06} in the variables $\msf{r}$, $\Phi$, and $\phi$ but only first-order accurate in $F$ and $f$ \cite{kreienbuehl:11b}. An explanation for the lower accuracy of $F$ and $f$ is that they are derived not only from $\msf{r}$ and $\Phi$ but also from $\phi$, which itself is derived from $\msf{r}$ and $\Phi$. That is, the inherited error can be expected to be much bigger for $F$ and $f$. According to \cite{thornburg:11}, an increased inherited error is a common feature of characteristic codes anyway. However, since our code reproduces Choptuik's classical findings, the lower accuracy of $F$ and $f$ is not a problem.

\section{\label{s:53}Black holes}

Since we are interested in the impact of various functions $\msf{q}$ on the black hole formation, we need to define a criterion that specifies black holes in characteristic coordinates $(\msf{u},\msf{v})$. With respect to the line element $-N^2d\msf{t}^2+A^2d\msf{r}^2+\msf{r}^2d\Omega^2$ such a criterion can be formulated with the help of the null expansion
\begin{equation*}
	\theta_+=\frac{\sqrt{2}}{\msf{r}A}
\end{equation*}
of a sphere of radius $\msf{r}$ embedded in a slice of constant coordinate time $\msf{t}$. Namely, a local minimum $\theta_+^{\mrm{min}}$ that is such that $\theta_+^{\mrm{min}}\to0$ as $\msf{t}$ increases, indicates the presence of an apparent horizon and, thus, the presence of a black hole. Note that in the coordinates $(\msf{t},\msf{r})$ we have no access to the expansion $\theta_-$ of a congruence of infalling null geodesics \cite{bardeen:83,choptuik:86,alcubierre:08}.\footnote{This is also true for double null and characteristic coordinates.}

Since $\theta_+$ is a scalar, the corresponding expression in characteristic coordinates $(\msf{u},\msf{v})$ is simply
\begin{equation*}
	\theta_+=\frac{1}{\msf{r}}\sqrt{\frac{2f}{F}}.
\end{equation*}
A good practical criterion for black hole formation in our model of deformed GR would therefore be to look for coordinates $(\msf{u}_{\mrm{AH}},\msf{v}_{\mrm{AH}})$ for which $\theta_+(\msf{u}_{\mrm{AH}},\msf{v}_{\mrm{AH}})$ is smaller than some user-defined threshold $\Theta_+\in\mbb{R}_+$. The code could then be stopped when $\theta_+<\Theta_+$ and the mass $\msf{M}_{\mrm{AH}}\equiv\msf{r}_{\mrm{AH}}/2\equiv\msf{r}(\msf{u}_{\mrm{AH}},\msf{v}_{\mrm{AH}})/2$ of the apparent horizon computed.\footnote{In $(\msf{t},\msf{r})$-coordinates one sets $A^2\equiv(1-2\msf{M}/\msf{r})^{-1}$ and black hole formation is signaled by $A\to\infty$, that is $\msf{M}\to\msf{r}/2$.} However, since in characteristic coordinates (and in double null coordinates as well) the code terminates ``by itself'' when an apparent horizon forms (the coordinates become singular) there is no need for the criterion $\theta_+<\Theta_+$. We simply keep a record of the values of $\theta_+^{\mrm{min}}$ and use the last one before the termination to get $\msf{M}_{\mrm{AH}}$.

For $\msf{q}=\msf{r}$ the characteristics follow the null geodesics and the mass $\msf{M}_{\mrm{AH}}$ is identical to the black hole mass $\msf{M}_{\mrm{BH}}$. However, at points where $\msf{q}<\msf{r}$ the slices $\msf{u}\in\mbb{R}$ are spacelike and they are timelike where $\msf{q}>\msf{r}$. It is then not clear from first principles whether we can set $\msf{M}_{\mrm{AH}}=\msf{M}_{\mrm{BH}}$. It is interesting that for $\msf{q}\neq\msf{r}$ the apparent horizons tend to form in the asymptotic region $\msf{v}\gg1$ of the computational domain but, unfortunately, this does not mean that they can be expected to coincide with the event horizon. The reason for this is that $\msf{r}$ remains small there, that is $\msf{r}\lesssim1$, which is exactly where the modifications are active. To address this issue, that is to understand the relation between $\msf{M}_{\mrm{AH}}$ and $\msf{M}_{\mrm{BH}}$ for $\msf{q}\neq\msf{r}$, we should really solve the geodesic equation over the entire spacetime (see \cite{ayal:97} in this context). On the one hand, it is unfortunate that this cannot be done in the given coordinates. On the other hand, we do not have to be too concerned about the causality implied by space- and timelike slices as the results we get for $\msf{M}_{\mrm{AH}}$ are very robust in that they appear insensitive to the choice of $\msf{q}$ (see the various plots below and, in particular, Section \ref{s:59}).

\section{\label{s:54}Questions of interest}

The questions we are interested in refer to Choptuik's mass scaling law \cite{choptuik:93,gundlach:07}. As far as the parameter $\phi_0$ is concerned, it reads
\begin{equation}\label{e:606}
	\msf{M}_{\mrm{AH}}\propto(\phi_0-\phi_0^*)^{\gamma},
\end{equation}
where $\phi_0^*$ is a critical amplitude in the sense that initial data with $\phi_0>\phi_0^*$ leads to a black hole, whereas smaller values do not. The value of the exponent $\gamma$ in \eref{e:606} is approximately $0.374$ \cite{gundlach:97} and Choptuik provided evidence for it to be universal. That is, the same value can be found for all members of a family of initial data profiles such as the members \eref{e:609} and \eref{e:611} of the family of massless scalar field profiles. Furthermore, Choptuik showed that for \eref{e:609} (with $\msf{v}^3/(1+\msf{v}^3)$ replaced by $\msf{v}^3$) there is a critical value $\msf{w}^*$ for which \eref{e:606} reads
\begin{equation}\label{e:612}
	\msf{M}_{\mrm{AH}}\propto(\msf{w}^*-\msf{w})^{\gamma}.
\end{equation}
The exponent $\gamma$ can once again be approximated by $0.374$.

The form of \eref{e:606} and \eref{e:612} implies that there is a naked singularity with $\msf{r}_{\mrm{AH}}=0$ corresponding to the parameters $\phi_0^*$ and $\msf{w}^*$. It can therefore be expected that quantum gravity corrections change the relations \eref{e:606} and \eref{e:612}. Indeed, this expectation was given more credibility by the work in \cite{husain:09,ziprick:09a}, where it was shown that similar corrections to those we consider lead to a mass gap. This means that a finite mass is required to form a black hole, which suggests that the naked singularity can no longer be found.

The questions we are interested in are as follows.
\begin{enumerate}[$1)$]
	\item Do \eref{e:606} and \eref{e:612} remain valid in the presence of various functions $\msf{q}$?
	\item Do the profiles given in \eref{e:609} and \eref{e:611} lead to different results?
	\item Is there a mass gap?
\end{enumerate}
Another interesting point concerns the oscillations that should be incorporated in the mass scaling laws \eref{e:606} and \eref{e:612} \cite{hod:97,gundlach:97}. The following question refers to these oscillations.
\begin{enumerate}[$4)$]
	\item What happens to the \tit{fine structure} of \eref{e:606} and \eref{e:612} in the presence of various functions $\msf{q}$?
\end{enumerate}

We will first address the questions $1)$, $2)$, and $3)$, and then turn our attention to question $4)$.

\section{\label{s:100}Results}

The figures in the remainder of this paper are the result of simulations performed at a resolution $\triangle\equiv2^{-8}\simeq4\cdot10^{-3}$. Furthermore, in all simulations but the ones leading to the GR points, we choose a grid with $\msf{v}_{\mrm{max}}=400\pm100$ and an initial position $\msf{v}_0=80$, which on the initial slice is significantly larger than the extent of the modified domain $\msf{v}\leq2$. In the GR runs a smaller grid with $\msf{v}_{\mrm{max}}=240$ is used but because of our intent to make a comparison with the modified GR runs, we do not alter $\msf{v}_0$. 

With the exception of the computations needed to analyse the validity of \eref{e:612}, where the width $\msf{w}$ is varied and the amplitude $\phi_0$ is fixed, we set $\msf{w}=32$ and varied $\phi_0$. The implemented initial scalar field profile is always that given in \eref{e:609}, except for the simulations addressing the question $2)$ above, where it is the one in \eref{e:611}. Finally, in all cases the point on the very left-hand side agrees with the critical parameter ($\phi_0^*$ or $w^*$) by $14$ decimal places. For all of the following mass scaling plots the simulations have been executed on the Atlantic Computational Excellence Network.

\subsection{The mass scaling law}

\begin{figure}
\begin{center}
	\includegraphics{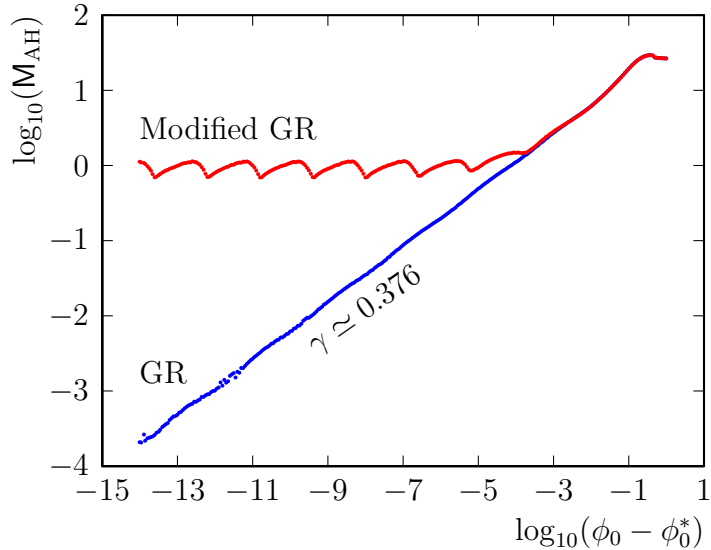}
	\caption[Modified mass scaling for $\msf{q}=\sqrt{1+\msf{r}^2}$.]{\label{f:12}An illustration of the impact of $\msf{q}=\sqrt{1+\msf{r}^2}$ on Choptuik's mass scaling law. Notice the mass gap, the oscillations, the overlap for $\log_{10}(\phi_0-\phi_0^*)\gtrsim-4$, and the scaling with respect to $\lambda$. The slope $\gamma\simeq0.376$ of the straight line about which the GR data oscillates was determined by a least squares fit over the interval $[-13,-3]$.}
\end{center}
\end{figure}

Figure \ref{f:12} shows that in the presence of $\msf{q}=\sqrt{1+r^2}$ Choptuik's mass scaling law is significantly modified.
The most obvious features of the new data points are a mass gap $\msf{M}_{\mrm{g}}$, an oscillatory behavior for $\log_{10}(\phi_0-\phi_0^*)\lesssim-4$, an overlap with the GR points for $\log_{10}(\phi_0-\phi_0^*)\gtrsim-4$, and a scaling with $\lambda$ (note that $\msf{M}_{\mrm{AH}}$ is denoting $M_{\mrm{AH}}/\lambda$). Finally, we will argue further down in favour of a certain universality and robustness of these features.

\subsubsection{Mass gap}

Since the critical solution is one of finite mass $\msf{M}_{\mrm{c}}\geq\msf{M}_{\mrm{g}}>0$ we are dealing with a type I phase transition in phase space. This is in contrast to the type II critical phenomena in GR \cite{gundlach:07}. It is interesting that the collapse of a massive scalar field in spherical symmetry does also exhibit type I phenomena \cite{brady:97}. However, unlike here, there continue to exist type II phase transitions for a sufficiently small scalar field mass. The fact that for a massive scalar field there are both type I and type II critical phenomena further strengthens the argument we presented in Section \ref{s:12}, according to which no scalar field potential can be added to the matter Lagrangian to reproduce the modified field equations we are solving.

\subsubsection{Oscillations}

We would like to emphasize that the oscillations are not a numerical artefact. This is because they show up for both initial scalar field profiles in \eref{e:609} and \eref{e:611}, as well as for all the functions $\msf{q}$ that we implemented (see Section \ref{s:59}). In the case of GR, the presence of oscillations in the mass scaling law is well known \cite{hod:97,gundlach:97} and its period in $\log_{10}$-$\log_{10}$ plots is approximated by $1.998$. The period of the GR data in Figure \ref{f:12} is about $2.010$ (we used an inverse discrete Fourier transformation over the interval $[-13,-3]$ on the horizontal axis). Further comments on the oscillations exhibited by the modified data are made in the concluding Section \ref{s:56}.

\subsubsection{Overlap}

The fact that the points from the modified collapse agree with those from the classical collapse for a range of ``large'' amplitudes was to be expected. It confirms the intuition that energetic initial data configurations behave as if the spacetime is classical; the effective quantum corrections cannot be ``seen''. Note that in Figure \ref{f:12} there are no points to the right of $\log_{10}(\phi_0-\phi_0^*)\simeq0$ as this is where the initial data is so ``heavy'' that it basically contains an apparent horizon.

\subsubsection{Scaling with $\lambda$}

Since $\msf{M}_{\mrm{AH}}$ is dimensionless we can read off Figure \ref{f:12} the impact a change in $\lambda$ has. If we decrease it, the mass $M_{\mrm{AH}}\equiv\lambda\msf{M}_{\mrm{AH}}$ of the apparent horizon has to become smaller and the data points corresponding to $M_{\mrm{AH}}$ are translated down towards larger negative numbers. However, the GR points serve as fixed points in this translation (``attractors'') in the sense that the overlapping region becomes larger. For $\lambda\to0$ the interval over which the modified data displays a mass gap disappears. That is, the oscillations are ``stretched out'' and the data is in accordance with the classical mass scaling law. 

\subsubsection{\label{s:59}Universality and robustness}

In this section we provide evidence for the claim that the mass gap, the oscillations, the overlap, and the scaling with $\lambda$ are
\begin{enumerate}[$1)$]
	\item universal with respect to the initial scalar field profile and
	\item robust under a change of the deformation function $\msf{q}$.
\end{enumerate}

\paragraph{Universality}

As for $1)$, the plot of the classical and the modified mass scaling relation for the initial data specified in \eref{e:611} displays a behavior that is identical to that apparent from Figure \ref{f:12} \cite{kreienbuehl:11b}. To be more precise, Table \ref{t:2} 
\begin{table}
\begin{center}
\begin{tabular}{|c|c|c|c|c|}
	\hline
	Equation & \eref{e:609} & \eref{e:611} & \eref{e:612} \\ \hline
	Domain & $[-14,-6]$ & $[-14,-6]$ & $[-12,-4]$ \\ 
	$[\log_{10}(\msf{M}_{\mrm{AH}})]_{\mrm{min}}$ & -0.160 & -0.162 & -0.162 \\
	$\delta[\log_{10}(\phi_0-\phi_0^*)]$ & 1.342 & 1.342 & 1.342 \\
	$\delta[\log_{10}(\msf{M}_{\mrm{AH}})]$ & 0.216 & 0.219 & 0.217 \\ \hline
\end{tabular}
\caption[Universality of the modified collapse.]{\label{t:2}A comparison of the specifics of the modified collapse for different initial data \eref{e:611} and a variation of $\msf{w}$ (see \eref{e:612}) instead of $\phi_0$ (see \eref{e:609}).}
\end{center}
\end{table}
shows that the size of the mass gap (we measured it by the minimum value of $\log_{10}(\msf{M}_{\mrm{AH}})$, that is we determined $[\log_{10}(\msf{M}_{\mrm{AH}})]_{\mrm{min}}$) and the specifics of the oscillations (comprising the period $\delta[\log_{10}(\phi_0-\phi_0^*)]$, which we determined by an inverse discrete Fourier transformation, and the amplitude $\delta[\log_{10}(\msf{M}_{\mrm{AH}})]$) agree for both \eref{e:609} and \eref{e:611} over the given intervals. To extend this notion of universality, we varied $\msf{w}$ and fixed $\phi_0$ for the initial data \eref{e:609}, just like Choptuik did. Once again the features mentioned previously are present \cite{kreienbuehl:11b} and the specifics are the same. The details are given in Table \ref{t:2}.

\paragraph{Robustness}

\begin{figure}
\begin{center}
	\subfigure[\label{f:91}A family of functions $\msf{q}$.]{\includegraphics[scale=1]{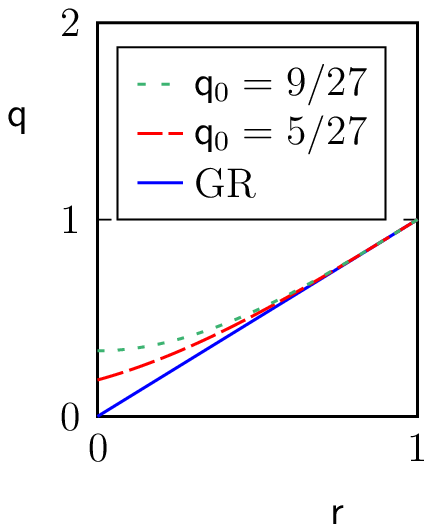}}\qquad
	\subfigure[\label{f:92}Mass scaling plots for various $\msf{q}_0$.]{\includegraphics[scale=1]{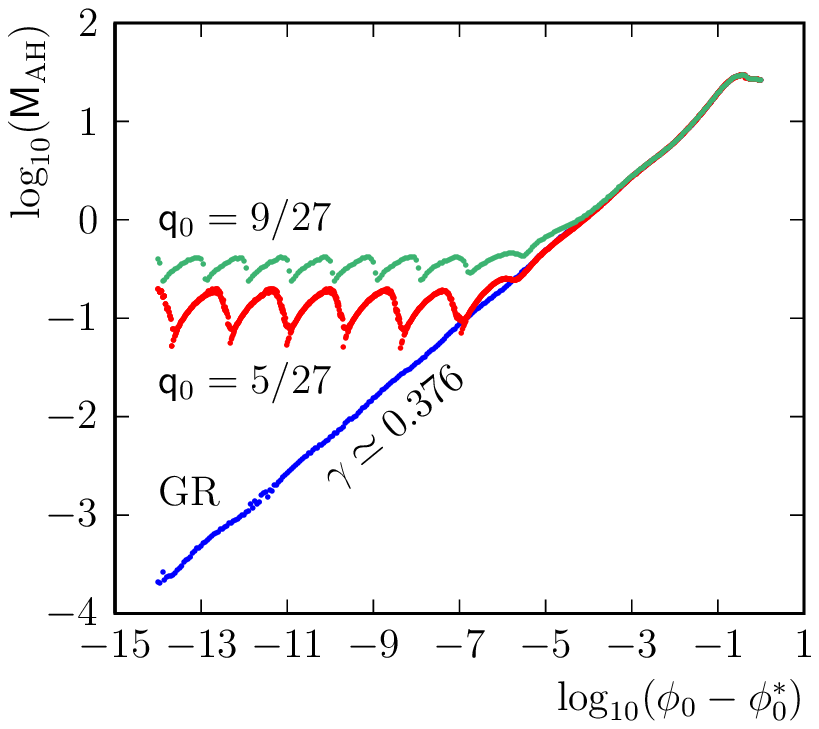}}
	\caption[A one-parameter family of functions $\msf{q}$.]{\label{f:9}A family of functions $\msf{q}$ parametrized by $\msf{q}_0$ and defined in \eref{e:614} is illustrated in Figure \ref{f:91}. Note that $\msf{q}_0$ has to be less than or equal to $1/3$ as otherwise the modified evolution equations are singular. The mass scaling plots are given in Figure \ref{f:92}. As compared to $\msf{q}_0=9/27$ and the GR case, there are five times more data points corresponding to $\msf{q}_0=5/27$ in order to better resolve the minima of the oscillations.}
\end{center}
\end{figure}

Regarding $2)$, we considered three different one-parameter families of functions $\msf{q}$. Two of them are discussed in \cite{kreienbuehl:11b} and Figure \ref{f:91} 
shows the function
\begin{equation}\label{e:614}
	\msf{q}\equiv\cases{
		\msf{q}_0+(1-3\msf{q}_0)\msf{r}+3\msf{q}_0\msf{r}^2-\msf{q}_0\msf{r}^3,&$\msf{r}<1$,\\
		\msf{r},&$\msf{r}\geq1$,
	}
\end{equation}
for two values of the parameter
\begin{equation*}
	\msf{q}_0\equiv\msf{q}(0).
\end{equation*}
For this function we need to impose the restriction $\msf{q}_0\leq1/3$ to guarantee that the modified evolution equations are not singular due to a local minimum of $\msf{q}$ (see \eref{e:551}). The function $\msf{q}$ in \eref{e:614} is interesting in that it reduces to $\msf{q}=\msf{r}$ in the limit $\msf{q}_0\to0$. Another aspect of \eref{e:614} is that it causes the characteristic directions $c_{\pm}$ to be spacelike for $\msf{r}<1$ and null otherwise. For $\msf{r}\geq1$ we can expect the causal structure of the spacetime to be identical to that implied by GR.

Figure \ref{f:92} shows that for two values of the $\msf{q}_0$ in \eref{e:614} the mass gap, the oscillations, the overlap, and the scaling with respect to $\lambda$ are present features. The conclusion is that Figure \ref{f:9} gives evidence in favour of the robustness proclaimed in $2)$.

\section{\label{s:56}Discussion}

In this paper we derived a set of effective equations for the massless scalar field in spherical symmetry. These equations deform GR but are nonetheless such that the constraint algebra closes. The deformation function $q$ can be interpreted as providing effective quantum gravity corrections based on an analogy with the inverse triad operator in LQG. Although this analogy is not exact, it is nevertheless of sufficient interest to numerically determine the resulting quantum effects on the gravitational collapse.

An interesting fact leading to the final form of our Equations \eref{e:550} and \eref{e:551} is that the characteristics of the scalar field do not coincide with the null directions of the metric. In the deformed theory, the characteristics can be timelike, null, or spacelike, depending on the form of the corrections. They are null if and only if there are no quantum corrections. This is the central reason why the equations in double null metric coordinates cannot be put into a useful form. However, when coordinates adapted to the scalar field characteristics are used, the equations simplify dramatically. 

A physical interpretation of the mismatch between the characteristics and the null directions of the metric is that the deformation can introduce violations of the dominant energy condition depending on the choice of the functions $Q_{(i)}$ and thus $q$. This can be seen directly by noting that we can define (in the notation of \eref{e:52}) an effective energy
density 
\begin{equation*}
	r^2A\varrho^{\mrm{e}}\equiv\frac{1}{Q_{(1)}}(Q_{(2)}\mca{H}_{\perp}^{(2)}+Q_{(3)}\mca{H}_{\perp}^{(3)})=\left(\frac{r}{q}\right)^2\mca{H}_{\perp}^{(2)}+\left(\frac{q}{r}\right)^2\mca{H}_{\perp}^{(3)}.
\end{equation*}
Then, since the diffeomorphism constraint is not modified, the dominant energy condition becomes
\begin{equation*}
	\varrho^{\mrm{e}}\geq\frac{|p_{\phi}\phi'|}{A}.
\end{equation*}
It appears that this inequality can be violated for a suitably chosen $q$. We can speculate that such a feature is common to all effective quantum gravity equations, however derived, because singularity avoidance without a violation of the dominant energy condition appears to be an unlikely occurrence. Our example provides one class of deformed and anomaly-free equations with this property. 

A violation of the dominant energy condition was also found by Ziprick and Kunstatter in \cite{ziprick:10}. They derive modified constraints from the dilaton action formulation of ``our'' model, because of which the corresponding Dirac algebra is necessarily anomaly-free. The question of how their approach relates to ours is an interesting one.

In this paper we also numerically solved and analysed the modified collapse equations that we derived in Section \ref{s:12}. The specific choice of $\msf{q}$ was guided by the conditions that its impact vanishes in the asymptotic region and that it regularizes the singular factors $1/\msf{r}^2$ in the gauge fixed Hamiltonian constraint. This is in agreement with the expectation that quantum gravity effects only modify regions of large energy densities. The same conditions for effective quantum gravity modifications for the gravitational collapse equations were also relied on in \cite{husain:09,ziprick:09a}. 

With the function $\msf{q}$ defined, we used a new combination of numerical methods to get an accurate code without any mesh refinement. That is, unlike in previous work \cite{garfinkle:94,husain:09}, we use a uniform grid in all our simulations. What is special about this grid is that it discretizes the dynamical fields not along the null lines of the metric but along the characteristic lines of the evolution equations. In GR they of course agree but because of the modifications the characteristics are better suited in our case.

The results of our analysis of the modified mass amplitude data are as follows. We found a mass gap, jagged oscillations, agreement with GR for sufficiently large amplitudes $\phi_0$, scaling with respect to the quantum length scale $\lambda$, a universality pertaining to different initial data profiles, and a robustness under a change of the function $\msf{q}$. What is interesting are the form of the oscillations and the robustness. The latter implies that the nature of the characteristic foliation defined by a particular $\msf{q}$ has no impact on the presence of the key findings but only on their fine structure.

As for the oscillations, they are new in comparison to earlier work \cite{husain:09} on quantum gravity corrected collapse in double null coordinates. The origin of the oscillations must be either that we use not exactly double null coordinates or that the way we implemented the modifications $\msf{q}$ does not lead to a violation of the Dirac algebra.

It is interesting that Ziprick and Kunstatter report in \cite{ziprick:09c} (see also \cite{taves:11} by Taves and Kunstatter) of similar oscillations resulting from the \tit{classical} collapse in Painlev\'e-Gullstrand coordinates. In \cite{ziprick:09a} they show that their quantum gravity modified model leads to the oscillations too. However, their analysis is based on the very different Painlev\'e-Gullstrand coordinates, which makes a comparison quite difficult. The Painlev\'e-Gullstrand coordinates allow for an evolution past the formation of apparent horizons, whereas our code has to terminate when such a horizon forms. We think that while the appearance of the oscillations in Ziprick and Kunstatter's work is in some way a result of the Painlev\'e-Gullstrand coordinates, \tit{the oscillations we find are more the result of the effective quantum gravity corrections $\msf{q}$ and the fact that they do not give rise to an anomaly in the Dirac algebra}.

A future project that is suggested by these considerations is a numerical analysis of our modified model in Painlev\'e-Gullstrand coordinates. This should clarify what the effect of the closure of the constraint algebra is. Furthermore, in Painlev\'e-Gullstrand coordinates it is also possible to determine the causal structure of the spacetime for various functions $\msf{q}$. In this context, an interesting question is to which extent our model can be compared to that of a collapsing $k$-essence \cite{akhoury:11,leonard:11}. Namely, it can be shown that the scalar field equation given in \eref{e:515} is equivalent to the equation $\tilde{\Box}\phi=0$, where $\tilde{\Box}\equiv\tilde{g}^{\mu\nu}\nabla_{\mu}\nabla_{\nu}$ is defined by the effective metric
\begin{equation*}
	d\tilde{\msf{s}}^2=-2\pde_{\msf{v}}\msf{r}Fd\msf{u}d\msf{v}+\msf{q}^2d\Omega^2.
\end{equation*}
On the one hand, this metric $\tilde{g}_{\mu\nu}$ implies that the scalar field evolves on a geometry, in which the smallest sphere has a surface area that is bounded from below by $4\pi(\min_{\msf{r}\geq0}\{\msf{q}(\msf{r})\})^2>0$. On the other hand, it defines a second causal structure that can be contrasted to the ``background'' structure given by $g_{\mu\nu}$. The latter will result in pictures that describe a $k$-essence. Another interesting question is what is happening to the self-similarity when the modifications are in action. To answer it, we can refine our code by an implementation of Garfinkle's algorithm \cite{garfinkle:94}. Finally, a related question that is still open is whether the critical solution of the modified collapse is an unstable star.

%\begin{acknowledgments}
\ack
This work was supported in part by the Natural Science and Engineering Research Council of Canada. We thank Martin Bojowald and Juan Reyes for providing a double null form of the scalar field-gravity equations in connection-triad variables, which motivated this work. We also thank Jack Gegenberg and Gabor Kunstatter for discussions, and the Atlantic Computational Excellence Network for computing resources.
%\end{acknowledgments}

\section*{References}

\bibliographystyle{unsrt}
\bibliography{/home/a/latex/biblio/b}

\begin{thebibliography}{10}

\bibitem{arnowitt:62}
R.~Arnowitt, S.~Deser, and C.~Misner.
\newblock {The Dynamics of General Relativity}.
\newblock In L.~Witten, editor, {\em {Gravitation: An Introduction to Current
  Research}}. Whiley, New York, 1962.

\bibitem{wheeler:68}
J.~A. Wheeler.
\newblock {Superspace and the Nature of Quantum Geometrodynamics}.
\newblock In C.~M. DeWitt and J.~A. Wheeler, editors, {\em {Battelles
  Rencontres: 1967 Lectures in Mathematics and Physics}}. Benjamin, New York,
  1968.

\bibitem{dewitt:67}
B.~S. DeWitt.
\newblock {Quantum theory of gravity. I. The canonical theory}.
\newblock {\em Phys. Rev.}, 160(5):1113--1148, 1967.

\bibitem{isham:91}
C.~J. Isham.
\newblock {Conceptual and geometrical problems in quantum gravity}.
\newblock {\em Lecture Notes in Physics}, 396:123--229, 1991.

\bibitem{kuchar:92}
K.~V. Kucha$\check{\trm{r}}$.
\newblock {Time and Interpretations of Quantum Gravity}.
\newblock In G.~Kunstatter, D.~Vincent, and J.~Williams, editors, {\em
  {Proceedings of the 4$^{th}$ Canadian Conference on General Relativity and
  Relativistic Astrophysics}}, pages 211--314. World Scientific, Singapore,
  1992.

\bibitem{isham:93}
C.~Isham.
\newblock {Canonical Quantum Gravity and the Problem of Time}.
\newblock In L.~A. Ibort and M.~A. Rodr\'iguez, editors, {\em {Integrable
  Systems, Quantum Groups and Quantum Field Theory}}, pages 157--287. Kluwer
  Academic, Dordrecht, Boston, 1993.

\bibitem{misner:72}
C.~W. Misner.
\newblock {Minisuperspace}.
\newblock In J.~Klauder, editor, {\em {Magic without Magic: John Archibald
  Wheeler}}, pages 441--473. W. H. Freeman, San Francisco, 1972.

\bibitem{bojowald:08b}
M.~Bojowlad.
\newblock {Loop quantum cosmology}.
\newblock {\em Living Rev. Relativity}, 11(4), 2008.
\newblock \url{http://www.livingreviews.org/lrr-2008-4}.

\bibitem{ashtekar:09}
A.~Ashtekar.
\newblock {Loop quantum cosmology: an overview}.
\newblock {\em Gen. Rel. Grav.}, 41(4):707--741, 2009.

\bibitem{choptuik:86}
M.~W. Choptuik.
\newblock {\em {A Study of Numerical Techniques for Radiative Problems in
  General Relativity}}.
\newblock PhD thesis, The University of British Columbia, 1986.

\bibitem{christodoulou:84}
D.~Christodoulou.
\newblock {Violation of cosmic censorship in the gravitational collapse of a
  dust cloud}.
\newblock {\em Commun. Math. Phys.}, 93:171--195, 1984.

\bibitem{goldwirth:87}
D.~S. Goldwirth and T.~Piran.
\newblock {Gravitational collapse of massless scalar field and cosmic
  censorship}.
\newblock {\em Phys. Rev. D}, 36(12):3575--3581, 1987.

\bibitem{christodoulou:86}
D.~Christodoulou.
\newblock {The problem of a self-gravitating scalar field}.
\newblock {\em Commun. Math. Phys.}, 105:337--361, 1986.

\bibitem{choptuik:92}
M.~W. Choptuik, D.~S. Goldwirth, and T.~Piran.
\newblock {A direct comparison of two codes in numerical relativity}.
\newblock {\em Class. Quantum Grav.}, 9(3):721--750, 1992.

\bibitem{choptuik:93}
M.~W. Choptuik.
\newblock Universality and scaling in gravitational collapse of a massless
  scalar field.
\newblock {\em Phys. Rev. Lett.}, 70(1):9--12, 1993.

\bibitem{garfinkle:94}
D.~Garfinkle.
\newblock {Choptuik scaling in null coordinates}.
\newblock {\em Phys. Rev. D}, 51(10):5558--5561, 1994.

\bibitem{husain:01}
V.~Husain and M.~Olivier.
\newblock {Scalar field collapse in three-dimensional AdS spacetime}.
\newblock {\em Class. Quantum Grav.}, 18(2):L1, 2001.

\bibitem{birukou:02}
M.~Birukou, V.~Husain, G.~Kunstatter, E.~Vaz, and M.~Olivier.
\newblock {Spherically symmetric scalar field collapse in any dimension}.
\newblock {\em Phys. Rev. Lett.}, 65(10):104036, 2002.

\bibitem{husain:03a}
V.~Husain, G.~Kunstatter, B.~Preston, and M.~Birukou.
\newblock {Anti-de Sitter gravitational collapse}.
\newblock {\em Class. Quantum Grav.}, 20(4):L23, 2003.

\bibitem{husain:09}
V.~Husain.
\newblock {Critical behavior in gravitational collapse}.
\newblock {\em Adv. Sci. Lett.}, 2(2):214--220, 2009.

\bibitem{husain:05b}
V.~Husain and O.~Winkler.
\newblock {Flat slice Hamiltonian formalism for dynamical black holes}.
\newblock {\em Phys. Rev. D}, 71(10):104001, 2005.

\bibitem{ziprick:09a}
J.~Ziprick and G.~Kunstatter.
\newblock {Dynamical singularity resolution in spherically symmetric black hole
  formation}.
\newblock {\em Phys. Rev. D}, 80(2):024032, 2009.

\bibitem{ziprick:10}
J.~Ziprick and G.~Kunstatter.
\newblock {Quantum corrected spherical collapse: a phenomenological framework}.
\newblock {\em arXiv:1004.0525v1}, 2010.

\bibitem{bojowald:06}
M.~Bojowald, M.~Kagan, P.~Singh, H.~H. Hernandez, and A.~Skirzewski.
\newblock {Hamiltonian cosmological perturbation theory with loop quantum
  gravity corrections}.
\newblock {\em Phys. Rev. D}, 74(12):123512, 2006.

\bibitem{bojowald:07}
M.~Bojowald and G.~M. Hossain.
\newblock {Loop quantum gravity corrections to gravitational wave dispersion}.
\newblock {\em Class. Quantum Grav.}, 24(18):4801, 2007.

\bibitem{bojowald:08a}
M.~Bojowald and G.~M. Hossain.
\newblock {Cosmological vector modes and quantum gravity effects}.
\newblock {\em Phys. Rev. D}, 77(2):023508, 2008.

\bibitem{reyes:09}
J.~D. Reyes.
\newblock {\em {Spherically Symmetric Loop Quantum Gravity: Connection to
  Two-Dimensional Models and Applications to Gravitational Collapse}}.
\newblock PhD thesis, The Pennsylvania State University, 2009.

\bibitem{berger:72}
B.~K. Berger, D.~M. Chitre, V.~E. Moncrief, and Y.~Nutku.
\newblock {Hamiltonian formulation of spherically symmetric gravitational
  fields}.
\newblock {\em Phys. Rev. D}, 5(10):2467--2470, 1972.

\bibitem{unruh:76}
W.~G. Unruh.
\newblock {Notes on black-hole evaporation}.
\newblock {\em Phys. Rev. D}, 14(4):870--892, 1976.

\bibitem{kuchar:94}
K.~V. Kuchar.
\newblock {Geometrodynamics of Schwarzschild black holes}.
\newblock {\em Phys. Rev. D}, 50(6):3961--3981, 1994.

\bibitem{ashtekar:87}
A.~Ashtekar.
\newblock {New Hamiltonian formulation of general relativity}.
\newblock {\em Phys. Rev. D}, 36(6):1587--1602, 1987.

\bibitem{barbero:95}
J.~F. Barbero.
\newblock {Real Ashtekar variables for {L}orentzian signature spacetimes}.
\newblock {\em Phys. Rev. D}, 51(10):5507--5510, 1995.

\bibitem{immirzi:97}
G.~Immirzi.
\newblock {Real and complex connections for canonical gravity}.
\newblock {\em Class. Quantum Grav.}, 14(10):L144, 1997.

\bibitem{thiemann:98a}
T.~Thiemann.
\newblock {Quantum spin dynamics}.
\newblock {\em Class. Quantum Grav.}, 15(4):839, 1998.

\bibitem{bojowald:08c}
M.~Bojowald, T.~Harada, and R.~Tibrewala.
\newblock {Lemaitre-Tolman-Bondi collapse from the perspective of loop quantum
  gravity}.
\newblock {\em Phys. Rev. D}, 78(6):064057, 2008.

\bibitem{bojowald:09}
M.~Bojowald, J.~D. Reyes, and R.~Tibrewala.
\newblock {Non-marginal LTB-like models with inverse triad corrections from
  loop quantum gravity}.
\newblock {\em Phys. Rev. D}, 80(8):084002, 2009.

\bibitem{husain:07a}
V.~Husain and O.~Winkler.
\newblock {Semiclassical states for quantum cosmology}.
\newblock {\em Phys. Rev. D}, 75(2):024014, 2007.

\bibitem{dirac:64}
P.~A.~M. Dirac.
\newblock {\em {Lectures on Quantum Mechanics}}.
\newblock Belfer Graduate School of Science, New York, 1964.

\bibitem{hanson:76}
A.~Hanson, T.~Regge, and C.~Teitelboim.
\newblock {\em {Constrained Hamiltonian Systems}}.
\newblock Accademia Nazionale dei Lincei, Roma, 1976.

\bibitem{henneaux:92}
M.~Henneaux and C.~Teitelboim.
\newblock {\em {Quantization of Gauge Systems}}.
\newblock Princeton University Press, Princeton, New Jersey, 1992.

\bibitem{gundlach:07}
J.~M. Martin-Garcia and C.~Gundlach.
\newblock {Critical phenomena in gravitational collapse}.
\newblock {\em Living Rev. Relativity}, 10(5), 2007.
\newblock \url{http://www.livingreviews.org/lrr-2007-5}.

\bibitem{berger:84}
M.~Berger and J.~Oliger.
\newblock {Adaptive mesh refinement of hyperbolic partial differential
  equations}.
\newblock {\em J. Comput. Phys.}, 53(3):484--512, 1984.

\bibitem{christodoulou:93}
D.~Christodoulou.
\newblock {Bounded variation solutions of the spherically symmetric
  {E}instein-scalar field equations}.
\newblock {\em Commun. Pure Appl. Math.}, XLVI:1131--1220, 1993.

\bibitem{strikwerda:04}
J.~C. Strikwerda.
\newblock {\em Finite Difference Schemes and Partial Differential Equations}.
\newblock Society for Industrial and Applied Mathematics, Philadelphia, 2nd
  edition, 2004.

\bibitem{husain:08}
V.~Husain and O.~Winkler.
\newblock {Quantum resolution of black hole singularities}.
\newblock {\em Class. Quantum Grav.}, 22:L127--L134, 2005.

\bibitem{halvorson:03}
H.~Halvorson.
\newblock {Complementarity of representations in quantum mechanics}.
\newblock {\em Stud. Hist. Philos. Sci. B Stud. Hist. Philos. Mod. Phys.},
  35(1):45--56, 2003.

\bibitem{ashtekar:03a}
A.~Ashtekar, S.~Fairhurst, and J.~L. Willis.
\newblock {Quantum gravity, shadow states, and quantum mechanics}.
\newblock {\em Class. Quantum Grav.}, 20(6):1031, 2003.

\bibitem{hoffman:01}
J.~D. Hoffman.
\newblock {\em {Numerical Methods for Engineers and Scientists}}.
\newblock CRC Press, New York, 2nd edition, 2001.

\bibitem{choptuik:91}
M.~W. Choptuik.
\newblock Consistency of finite-difference solutions of einstein's equations.
\newblock {\em Phys. Rev. D.}, 44(10):3124--3135, 1991.

\bibitem{pretorius:02}
F.~Pretorius.
\newblock {\em {Numerical Simulations of Gravitational Collapse}}.
\newblock PhD thesis, The University of British Columbia, 2002.

\bibitem{pretorius:06}
F.~Pretorius and M.~W. Choptuik.
\newblock {Adaptive mesh refinement for coupled elliptic-hyperbolic systems}.
\newblock {\em J. Comput. Phys.}, 218(1):246--274, 2006.

\bibitem{kreienbuehl:11b}
A.~Kreienbuehl.
\newblock {\em {Quantum Cosmology, Polymer Matter, and Modified Collapse}}.
\newblock PhD thesis, The University of New Brunswick, 2011.

\bibitem{thornburg:11}
J.~Thornburg.
\newblock {Adaptive mesh refinement for characteristic grids}.
\newblock {\em Gen. Relat. Gravit.}, 43(5):1211--1251, 2011.

\bibitem{bardeen:83}
J.~M. Bardeen and T.~Piran.
\newblock {General relativistic axisymmetric rotating systems: coordinates and
  equations}.
\newblock {\em Physics Reports (Review Section of Physics Letters)},
  96(4):205--250, 1983.

\bibitem{alcubierre:08}
M.~Alcubierre.
\newblock {\em {Introduction to 3+1 Numerical Relativity}}.
\newblock Oxford University Press, Oxford, 2008.

\bibitem{ayal:97}
S.~Ayal and T.~Piran.
\newblock {Spherical collapse of a massless scalar field with semiclassical
  corrections}.
\newblock {\em Phys. Rev. D}, 56(8):4768--4774, 1997.

\bibitem{gundlach:97}
C.~Gundlach.
\newblock {Understanding critical collapse of a scalar field}.
\newblock {\em Phys. Rev. D}, 55(2):695--713, 1997.

\bibitem{hod:97}
S.~Hod and T.~Piran.
\newblock {Fine structure of Choptuik's mass-scaling relation}.
\newblock {\em Phys. Rev. D}, 55(2):R440--R442, 1997.

\bibitem{brady:97}
P.~R. Brady, C.~M. Chambers, and S.~M. C.~V. Gon{\c{c}}alves.
\newblock {Phases of massive scalar field collapse}.
\newblock {\em Phys. Rev. D}, 56(10):R6057--R6061, 1997.

\bibitem{ziprick:09c}
J.~Ziprick and G.~Kunstatter.
\newblock {Numerical study of black-hole formation in Painlev{\'e}-Gullstrand
  coordinates}.
\newblock {\em Phys. Rev. D}, 79(10):101503(R), 2009.

\bibitem{taves:11}
T.~Taves and G.~Kunstatter.
\newblock {Higher dimensional Choptuik scaling in Painleve Gullstrand
  coordinates}, 1997.
\newblock arXiv:1105.0878v1.

\bibitem{akhoury:11}
R.~Akhoury, D.~Garfinkle, and R.~Saotome.
\newblock Gravitational collapse of {$k$}-essence, 2011.
\newblock arXiv:1103.0290v1.

\bibitem{leonard:11}
D.~Leonard, 2011.
\newblock talk at the Black Holes VIII conference.

\end{thebibliography}

\end{document}